\documentclass[12pt]{article}
\usepackage{graphicx}

\usepackage{epsf}
\def\beq{\begin{equation}}
\def\eeq{\end{equation}}
\def\bea{\begin{eqnarray}}
\def\eea{\end{eqnarray}}
\def\ve{\vert}
\def\vel{\left|}
\def\ver{\right|}
\def\nnb{\nonumber}
\def\ga{\left(}
\def\dr{\right)}
\def\aga{\left\{}
\def\adr{\right\}}
\def\rar{\rightarrow}
\def\nnb{\nonumber}
\def\la{\langle}
\def\ra{\rangle}
\def\ba{\begin{array}}
\def\ea{\end{array}}
\def\bea{\begin{eqnarray}}
\def\eea{\end{eqnarray}}

\def\ds{\displaystyle}
\def\ve{\vert}
\def\vel{\left|}
\def\ver{\right|}
\def\nnb{\nonumber}
\def\ga{\left(}
\def\dr{\right)}
\def\aga{\left\{}
\def\adr{\right\}}
\def\rar{\rightarrow}
\def\nnb{\nonumber}
\def\la{\langle}
\def\ra{\rangle}
\def\lla{\left<}
\def\rra{\right>}
\def\simlt{\stackrel{<}{{}_\sim}}
\def\simgt{\stackrel{>}{{}_\sim}}
\begin{document}
\def\beq{\begin{equation}}
\def\eeq{\end{equation}}
\def\bea{\begin{eqnarray}}
\def\eea{\end{eqnarray}}
\def\ve{\vert}
\def\vel{\left|}
\def\ver{\right|}
\def\nnb{\nonumber}
\def\ga{\left(}
\def\dr{\right)}
\def\aga{\left\{}
\def\adr{\right\}}
\def\rar{\rightarrow}
\def\nnb{\nonumber}
\def\la{\langle}
\def\ra{\rangle}
\def\lla{\left<}
\def\rra{\right>}
\def\ba{\begin{array}}
\def\ea{\end{array}}
\def\BcDll{$B_c \rar D_s \ell^+ \ell^-$}
\def\BcDllmu{$B_c \rar D_s \mu^+ \mu^-$}
\def\BcDlltau{$B_c \rar D_s \tau^+ \tau^-$}
\def\tepm{$B \rar K \mu^+ \mu^-$}
\def\tept{$B \rar K \tau^+ \tau^-$}
\def\ds{\displaystyle}



\newskip\humongous \humongous=0pt plus 1000pt minus 1000pt
\def\caja{\mathsurround=0pt}
\def\eqalign#1{\,\vcenter{\openup1\jot
\caja   \ialign{\strut \hfil$\displaystyle{##}$&$
\displaystyle{{}##}$\hfil\crcr#1\crcr}}\,}


\def\simlt{\stackrel{<}{{}_\sim}}
\def\simgt{\stackrel{>}{{}_\sim}}



\def\bos{\lower 0.5cm\hbox{{\vrule width 0pt height 1.2cm}}}
\def\boss{\lower 0.35cm\hbox{{\vrule width 0pt height 1.cm}}}
\def\aaa{\lower 0.cm\hbox{{\vrule width 0pt height .7cm}}}
\def\dol{\lower 0.4cm\hbox{{\vrule width 0pt height .5cm}}}


\title{ {\Large {\bf
Rare \BcDll decay beyond the standard model } } }

\author
{{\small  U. O. Yilmaz \thanks {e-mail: uoyilmaz@mersin.edu.tr}}\\
{\small Physics Department, Mersin University 33343
Ciftlikkoy Mersin, Turkey} }
\date { }

\begin{titlepage}
\maketitle
\thispagestyle{empty}

\begin{abstract}
The rare \BcDll decay is investigated by
using the most general model independent effective Hamiltonian. The
general expressions of longitudinal, normal and
transversal polarization asymmetries for $\ell^-$  and $\ell^+$ and the combined
asymmetries of them are found. The dependencies of the branching ratios and polarizations on
the new Wilson coefficients
are presented.
The analysis shows that the branching ratios and the lepton polarization
asymmetries are very sensitive to the scalar and tensor type
interactions. These results will be very useful in searching new physics
beyond the standard model.
\end{abstract}

\end{titlepage}

\section{Introduction}
Investigation of rare $B$ meson decays, induced by flavor--changing
neutral current (FCNC) $b \rar s,d$ \cite{Ali1} transitions, is an
important source of new physics. These transitions take place at
loop level in the SM so they can be used to test the gauge structure
of the SM and provide a suitable tool of looking for new physics. In
rare B meson decays, new physics contributions may appear in two
different ways; modifying  Wilson coefficients in the SM or adding
new structures in the SM effective Hamiltonian.

Since the CLEO observation of $b \rar s \, \gamma$ process
\cite{CLEO}, decays of $B_{u,d,s}$ mesons have been the subject of
many investigations.  These studies will be even more complete if
similar studies for $B_c$, discovered by CDF Collaboration
\cite{CDF}, are also included.

In the mean time, the study of the $B_c$ meson is by itself quite
interesting too, since it has some outstanding features
\cite{Colangelo}--\cite{Ivanov2} . It is the lowest bound state of
two heavy quarks ($b$ and $c$) with explicit flavor that can be
compared with the charmonium ($c\bar{c}$- bound state) and bottomium
($b\bar{b}$- bound state) which have implicit flavor. The
implicit-flavor states decay strongly and electromagnetically
whereas the $B_c$ meson decays weakly. The major difference between
the weak decay properties of $B_c$ and $B_{u,d,s}$ is that those of
the latter ones are described very well in the framework of the
heavy quark limit, which gives some relations between the form
factors of the physical process. In case of $B_c$ meson, the heavy
flavor and spin symmetries must be reconsidered because both $b$ and
$c$ are heavy.

On the experimental side, like the running B factories in KEK and
SLAC, also encourages the study of the rare B meson decays and most
of the rare $B_c$ decays are believed to be accessible in future
experiments at hadronic colliders, such as the LHC-B. The Tevatron
experiments see around 100 $B_c$ semileptonic decays and a
luminosity upgrade by a factor of ten is discussed for 2013. Atlas
and CMS experiments will have more $B_c$ but there will be more
background. This scene may not be hopeful for understanding $B_c$.
On the other side, rapid progress on experimental techniques are
still encouraging.

Measurement of the lepton polarization is an efficient way in
establishing the new physics beyond the SM
\cite{Kruger}--\cite{uoyilmaz}. In this work we present a  study of the
branching ratio and lepton polarizations in the exclusive \BcDll
 $~(\ell = \mu,~\tau)$ decay for a general form of the
effective Hamiltonian including all possible form of interactions
in a model independent way  without forcing concrete values for
the Wilson coefficients corresponding to any specific model. To
make predictions about such an  exclusive decay, one requires the
additional knowledge about form factors, i.e., the matrix elements
of the effective Hamiltonian between initial and final meson
states.  This problem, being a part of the nonperturbative sector
of  QCD, lacks a precise solution. In literature there are a
number of  different approaches to calculate the decay form
factors of \BcDll decay
 such as light front, constituent quark models,
and a relativistic quark model proposed in \cite{Faessler}.
In this work we will use the weak decay form factors calculated in \cite{Faessler}.

The work is organized as follows. In section 2, we derive the matrix
element starting from the effective Hamiltonian for the quark level
process and using the appropriate form factors. Then, we present the
model independent expressions for the longitudinal, transversal and
normal polarizations of leptons and combined lepton-antilepton
asymmetries. We give our numerical results and discussion in section
3.

\section{Effective Hamiltonian and Lepton Polarizations}

The \BcDll decay is described at the quark level by the $b \rar  s
\ell^+ \ell^- $ transition in the standard effective Hamiltonian
approach. This Hamiltonian includes all possible terms calculated
independent of any models. The effective Hamiltonian for this
process can be written in terms of twelve model independent
four-Fermi interactions, as follows\cite{Fukae}: \bea \label{effH}
{\cal H}_{eff} &=& \frac{G\alpha}{\sqrt{2} \pi}
     V_{ts}V_{tb}^\ast
     \Bigg\{ C_{SL} \, \bar s i \sigma_{\mu\nu} \frac{q^\nu}{q^2}\, L \,b
    \, \bar \ell \gamma^\mu \ell + C_{BR}\, \bar s i \sigma_{\mu\nu}
    \frac{q^\nu}{q^2} \,R\, b \, \bar \ell \gamma^\mu \ell \nnb \\
&+&C_{LL}^{tot}\, \bar s_L \gamma_\mu b_L \,\bar \ell_L \gamma^\mu \ell_L +
    C_{LR}^{tot} \,\bar s_L \gamma_\mu b_L \, \bar \ell_R \gamma^\mu \ell_R +
    C_{RL} \,\bar s_R \gamma_\mu b_R \,\bar \ell_L \gamma^\mu \ell_L \nnb \\
&+&C_{RR} \,\bar s_R \gamma_\mu b_R \, \bar \ell_R \gamma^\mu \ell_R +
    C_{LRLR} \, \bar s_L b_R \,\bar \ell_L \ell_R +
    C_{RLLR} \,\bar s_R b_L \,\bar \ell_L \ell_R \\
&+&C_{LRRL} \,\bar s_L b_R \,\bar \ell_R \ell_L +
    C_{RLRL} \,\bar s_R b_L \,\bar \ell_R \ell_L+
    C_T\, \bar s \sigma_{\mu\nu} b \,\bar \ell \sigma^{\mu\nu}\ell \nnb \\
&+&i C_{TE}\,\epsilon^{\mu\nu\alpha\beta} \bar s \sigma_{\mu\nu} b \,
    \bar \ell \sigma_{\alpha\beta} \ell  \Bigg\}~, \nnb
\eea
where $L={1-\gamma_5}/{2}$ and $R={1+\gamma_5}/{2}$ are the chiral projection operators
and $C_X$ are the coefficients of the four--Fermi interactions. The coefficients $C_{SL}$ and $C_{BR}$ are the nonlocal Fermi
interactions corresponding to $-2 m_s C_7^{eff}$ and $-2 m_b C_7^{eff}$
in the SM, respectively. The $C_{LL}$, $C_{LR}$, $C_{RL}$ and $C_{RR}$ terms are the vector type interactions, two of which
are vector interactions containing $C_{LL}^{tot}$ and $C_{LR}^{tot}$ do already exist in the SM
in combinations of the form $(C_9^{eff}-C_{10})$ and $(C_9^{eff}+C_{10})$.
Therefore, we write
\bea
C_{LL}^{tot} &=& C_9^{eff} - C_{10} + C_{LL}~, \nnb \\
C_{LR}^{tot} &=& C_9^{eff} + C_{10} + C_{LR}~, \nnb
\eea
so that that $C_{LL}^{tot}$ and $C_{LR}^{tot}$ describe the
sum of the contributions from SM and new physics. The terms with
coefficients $C_{LRLR}$, $C_{RLLR}$, $C_{LRRL}$ and $C_{RLRL}$ describe
the scalar type interactions and the last two terms, $C_T$ and $C_{TE}$, describe the tensor type
interactions.

Having the general form of four--Fermi interaction for the $b \rar s
\ell^+ \ell^-$ transition, the next task is to calculate  the
matrix element for the \BcDll decay which can be expressed in terms of the invariant form factors over $B_c$ and $D_s$.
These form factors are weak decay form factors \cite{Faessler}.
\bea
\label{ilk}
    \lla D_s(p_{D_{s}}) \vel \bar s  i \sigma_{\mu \nu} q^{\nu} b \ver B_c(p_{B_c}) \rra
    &=&- \frac {F_{T}} {m_{B_c} + m_{D_{s}}} \left [ {(P_{B_c} +
    P_{D_{s}})}_\mu q^2 - q_{\mu} (m^{2}_{B_c} - m^{2}_{D_{s}})\right],  \\
\label{iki}
    \lla D_s(p_{D_{s}}) \vel \bar s  \gamma _{\mu} b \ver B_s(p_{B_c}) \rra
    &=& F_{+}(q^2) {(P_{B_c} + P_{D_{s}})}_{\mu} + F_{-}(q^2) q_{\mu}~,  \\
\label{ucc}
    \lla D_s(p_{D_{s}}) \vel \bar s  b \ver B_s(p_{B_c}) \rra
    &=& F_{-} \frac {q^2} {m_b - m_s} + F_{+} \frac {m^2_{B_c} -
    m^2_{D_{s}}} {m_b - m_s}~,
\eea
and
\bea
\label{dort}
    \lla D_s(p_{D_{s}}) \vel \bar s \sigma_{\mu \nu} b \ver B_s(p_{B_c}) \rra
    = i \frac{F_{T}} {m_{B_c} + m_{D_{s}}} \left[{(P_{B_c} +
    P_{D_{s}})}_{\mu} q_{\nu} - q_{\mu} {(P_{B_c} + P_{D_{s}})}_{\nu}
    \right]~,
\eea
where $q=p_{B_c}-p_{D_s}$ is the momentum transfer.

We can write the matrix element of the \BcDll decay using Eq. (\ref {effH})-(\ref {dort}) as
\bea
 \label{had}
    {\cal M}(B_c\rightarrow D_s \ell^{+}\ell^{-}) &=&
    \frac{G \alpha}{4 \sqrt{2} \pi} V_{tb} V_{ts}^\ast
 \Bigg\{
    \bar \ell \gamma^\mu \ell \, \Big[
    A(P_{B_c} + P_{D_{s}})_\mu +Bq_\mu \Big] \nnb \\
&&+ \bar \ell \gamma^\mu \gamma_5 \ell \, \Big[
    C(P_{B_c} + P_{D_{s}})_\mu +Dq_\mu \Big]
    + \bar \ell \ell Q +\bar \ell \gamma_5 \ell N \nnb \\
&&+4 \bar \ell \sigma^{\mu \nu} \ell ( i G) \Big[(P_{B_c} + P_{D_{s}})_\mu
    q_\nu - (P_{B_c} + P_{D_{s}})_\nu +q_\mu \Big] \nnb \\
&&+4 \bar \ell \sigma_{\alpha\beta}  \ell \epsilon^{\mu \nu \alpha
    \beta} H \Big[(P_{B_c} + P_{D_s})_\mu q_\nu -(P_{B_c} + P_{D_{s}})_\nu q_\mu
    \Big]
    \Bigg\}~,
\eea
where
\bea
\label{as} A &=& (C_{LL}^{tot} + C_{LR}^{tot} +
    C_{RL} + C_{RR}) F_{+}(q^2) - 2(C_{SL}) + C_{BR})
    \frac{F_T}{m_{B_c}+m_{D_s}} ~, \nnb \\
B&=&(C_{LL}^{tot} + C_{RL}^{tot} + C_{RL} + C_{RR}) F_{-}(q^2) +
    2 (C_{SL}+ C_{BR})F_{T} \frac {(m_{B_c}-m_{D_s})} {q^2} ~, \nnb \\
C &=& (C_{LR}^{tot} + C_{RR} - C_{LL}^{tot} - C_{RL})F_{+}(q^2)  ~,
    \nnb \\
D &=&  (C_{LR}^{tot} + C_{RR} - C_{LL}^{tot} - C_{RL})F_{-}(q^2)  ~,
    \nnb \\
Q &=& (C_{LRLR} + C_{RLLR} +C_{LRRL} + C_{RLRL})
    \Big[ F_{-} \frac {q^2} {m_b - m_s}
        + F_{+} \frac {m^2_{B_c} - m^2_{D_s}} {m_b - m_s} \Big] ~, \nnb \\
N &=&  (C_{LRLR} + C_{RLLR} - C_{LRRL} - C_{RLRL} )
    \Big[F_{-} \frac {q^2} {m_b - m_s}
        + F_{+} \frac {m^2_{B_c} - m^2_{D_s}} {m_b - m_s} \Big] ~, \nnb \\
G &=& - C_{T} \frac {F_{T}} {m_{B_c} + m_{D_s} }  ~, \nnb \\
H &=& -C_{TE} \frac {F_{T}} {m_{B_c} + m_{D_s} }~.
\eea
In order to calculate the final lepton polarizations, we define the orthogonal unit vector $S_{i}^{- \mu}$ in the rest frame of $\ell^-$ and
$S_{i}^{+ \mu}$ in the rest frame of $\ell^+$ and the polarization of the leptons along
the longitudinal ($L$), transversal ($T$) and normal ($N$)
directions, as done before \cite{Kruger, Fukae}, by
\bea
\label{pol} S_L^{-\mu} &\equiv& (0,\vec{e}_L^{\,-})
=
\ga 0,\frac{\vec{p}_-}{\vel \vec{p}_- \ver} \dr~, \nnb \\
S_N^{-\mu} &\equiv& (0,\vec{e}_N^{\,-}) =
\ga 0,\frac{\vec{p} \times \vec{p}_-}
{\vel \vec{p} \times \vec{p}_- \ver} \dr~, \nnb \\
S_T^{-\mu} &\equiv& (0,\vec{e}_T^{\,-}) =
\ga 0, \vec{e}_N^{\,-} \times \vec{e}_L^{\,-} \dr~, \\
S_L^{+\mu} &\equiv& (0,\vec{e}_L^{\,+}) =
\ga 0,\frac{\vec{p}_+}{\vel \vec{p}_+ \ver} \dr~, \nnb \\
S_N^{+\mu} &\equiv& (0,\vec{e}_N^{\,+}) =
\ga 0,\frac{\vec{p} \times \vec{p}_+}
{\vel \vec{p} \times \vec{p}_+ \ver} \dr~, \nnb \\
S_T^{+\mu} &\equiv& (0,\vec{e}_T^{\,+}) =
\ga 0, \vec{e}_N^{\,+} \times \vec{e}_L^{\,+} \dr~, \nnb
\eea
where $\vec{p}_\pm$ and $\vec{p}$ are the three momenta of $\ell^\pm$ and
$D_s$ meson in the center of mass (CM) frame of the lepton pair
system, respectively. The longitudinal unit vectors $S_L^-$ and $S_L^+$ are
boosted to the CM frame of $\ell^+ \ell^-$ by Lorentz transformation,
\bea
\label{bs}
S^{-\mu}_{L,\, CM} &=& \ga \frac{\vel \vec{p}_- \ver}{m_\ell},
\frac{E_\ell \,\vec{p}_-}{m_\ell \vel \vec{p}_- \ver} \dr~, \nnb \\
S^{+\mu}_{L,\, CM} &=& \ga \frac{\vel \vec{p}_- \ver}{m_\ell},
- \frac{E_\ell \, \vec{p}_-}{m_\ell \vel \vec{p}_- \ver} \dr~,
\eea
while vectors of perpendicular directions are not changed by boost.

As $\vec{n}^{\,\pm}$ being any spin direction of the $\ell^{\pm}$, in the rest frame of the leptons,
the differential decay rate of the \BcDll decay can be written in the following form:
\bea
\label{ddr}
\frac{d\Gamma(\vec{n}^{\pm})}{ds} = \frac{1}{2}
    \ga \frac{d\Gamma}{ds}\dr_0
        \Bigg[ 1 + \Bigg( P_L^{\pm} \vec{e}_L^{\,\pm} + P_N^{\pm}
            \vec{e}_N^{\,\pm} + P_T^{\pm} \vec{e}_T^{\,\pm} \Bigg) \cdot
                \vec{n}^{\pm} \Bigg].
\eea
Here, $s=q^2/m_{B_c}^2$, the superscripts $^+$ and $^-$ correspond to the $\ell^+$ and $\ell^-$ cases and
$(d\Gamma / ds)_{0}$ corresponds to the unpolarized decay
rate, whose explicit form is
\bea
\label{unp}
\ga \frac{d \Gamma}{ds}\dr_0 &=& \frac{G^2 \alpha^2 m_{B_c}}{2^{14} \pi^5 }
    \vel V_{tb} V_{ts}^\ast \ver^2 \sqrt{\lambda} v \Delta
\eea
where
\bea
\label{delta}
\Delta &=& \frac {1024} {3} s v^2 \lambda m_{B_c}^6 \vel H \ver^2 +
         \frac{256} {3} s (3-2v^2) \lambda m_{B_c}^6 \vel G \ver^2
         - \frac {4} {3} (v^2-3) \lambda m_{B_c}^4 \vel A \ver \nnb \\
&-& 128 m_{\ell} \lambda m_{B_c}^4 Re(AG^*)
        +  32 (1-r) m_{\ell}^2 m_{B_c}^2 Re(CD^*)
            + 16 s m_{\ell} m_{B_c}^2 \vel D \ver^2 \nnb \\
    &+& 4 s m_{B_c}^2 \vel N \ver^2
        + \frac {4} {3} m_{B_c}^4 s [2 \lambda - (1- v^2)(2 \lambda - 3(1-r)^2)] \vel C \ver^2 \nnb \\
&+& 16 (1-r) m_{\ell} m_{B_c}^2 Re(CN^*)
    + 16 s m_{\ell} m_{B_c}^2 Re(DN^*)
       + 4 s v^2 m_{B_c}^2 \vel Q \ver^2
\eea
and $\lambda= 1 + r^2 + s^2 -2r-2s-2rs$, $r=m_{D_s}^2/m_{B_c}^2$ and lepton velocity is $v=\sqrt{1-{4m_\ell^2}/{q^2}}$.

The polarizations $P_L^{\pm}$, $P_T^{\pm}$ and $P_N^{\pm}$ in Eq. (\ref{ddr}) are defined by
\bea
P_i^{\pm}(q^2) = \frac{\ds{\frac{d \Gamma}{dq^2}
                   (\vec{n}^{\pm}=\vec{e}_i^{\,\pm}) -
                   \frac{d \Gamma}{dq^2}
                   (\vec{n}^{\pm}=-\vec{e}_i^{\,\pm})}}
              {\ds{\frac{d \Gamma}{dq^2}
                   (\vec{n}^{\pm}=\vec{e}_i^{\,\pm}) +
                  \frac{d \Gamma}{dq^2}
                  (\vec{n}^{\pm}=-\vec{e}_i^{\,\pm})}}~, \nnb
\eea
for $i=L,~N,~T$. Here, $P_L^{\pm}$ and $P_T^{\pm}$ represent the longitudinal and transversal asymmetries of the charged lepton $\ell^{\pm}$
in the decay plane and $P_N^{\pm}$ is the normal component to both of them.
After calculations, the longitudinal
polarization of the $\ell^{\pm}$ is
%
 \bea
 \label{plm} P_{L}^\pm = \frac {8 m_{B_c}^2 v} {\Delta} \Big[
             \mp \frac{2} {3} m_{B_c}^2 \lambda Re(AC^*)
                \pm \frac {16} {3} m_{B_c}^2 m_{\ell} \lambda Re(CG^*)
                -\frac {32} {3} m_{B_c}^2 m_\ell \lambda Re(AH^*) \nnb \\
   - 2 m_\ell (1-r) Re(CQ^*)
         + \frac {128} {3} m_{B_c}^4 \lambda s Re(GH^*)
         - 2 m_\ell s Re(DQ^*)
         - s Re(NQ^*) \Big]
 \eea
where $\Delta$ is given in Eq. (\ref{delta}).

In a similar way, we find the transverse polarization $P_T^{\pm}$
%
\bea
\label{ptm}
    P_{T}^\pm = \frac{2 m_{B_c}^3 \pi \sqrt{s \lambda}} {\Delta} \Big[
         \pm \frac{2} {s} m_\ell (1-r) Re(AC^*)
         \pm \frac {32}  {s} m_{\ell}^2 (1-r) Re(CG^*)
         \mp 2 m_\ell Re(AD^*) \nnb \\
    \pm 32 m_{\ell}^2 Re(DG^*)
         \mp  Re(AN^*)
        \pm 16 m_\ell Re(GN^*)
         + v^2 Re(CQ^*)
    \Big],
\eea In the limit of $m_{\ell} \rar0$, the transverse polarization
is due the scalar terms. This can give new information about new
physics.

Finally, the normal polarization $P_N^{\pm}$ is given by
\bea
\label{pnm}
    P_{N}^\pm = \frac{2 m_{B_c}^3 \pi v \sqrt{s \lambda}}{\Delta}\Big[
        - 2 m_\ell Im(CD^*)
            -  Im(CN^*)
                \pm Im(AQ^*)
                   \mp 16 m_\ell Im(GQ^*) \Big].
\eea
In this work we assume all form factors and all new Wilson
coefficients are real. Therefore, the only contribution to
$P_{N}^\pm$ in Eq. (\ref {pnm}) comes from $\pm Im(AQ)$ term since
only the function A has an imaginary part coming from $C_{9}^{eff}$.
For this reason, $P_{N}^- = P_{N}^+ = 0$ in the SM and scalar terms
in $Q$ makes normal polarization nonzero beyond the SM. This
observable result gives useful clue about new physics.

Since in the SM $P_L^-+P_L^+=0$, $P_T^- - P_T^+ \approx 0$ and
$P_N^-+P_N^+= 0$, combined analysis of the lepton and antilepton
polarizations can be another useful source of new physics \cite
{Fukae}.

Using Eq. (\ref{plm}) we get combined longitudinal polarization
\bea
\label{lpl}
P_L^- + P_L^+ = \frac {16 m^2_{B_c} v} {3 \Delta}
    \Big [ 128 s \lambda m^4_{B_c} Re(GH^*) -
        32 m_{\ell} m^2_{B_c}\lambda Re(AH^*) \nnb \\
  - 3 s Re(NQ^*) -
            6 (1-r) m_\ell Re(C Q^*)-
                6 m_\ell s Re(D Q^*) \Big],
\eea and combined transversal polarization is the difference of the
lepton and antilepton polarizations and can be calculated from Eq.
(\ref{ptm})
\bea
\label{tmt}
P_T^- - P_T^+ = \frac{4 m^3_{B_c}\pi \sqrt{s \lambda}} {\Delta}
\Big[ - \frac{32 m^2_{\ell} (1-r)} {s} Re(CG^*)
    + \frac{2 m_{\ell} (1-r)} {s} Re(AC^*)
     \nnb \\
+ 2 m_{\ell}  Re(AD^*)
    - 32 m^2_{\ell} Re(DG^*)
    + Re(AN^*)
    - 16 m_{\ell} Re(GN^*) \Big].
\eea
The terms containing the SM contribution to the $P_L^-+P_L^+$ in Eq.
(\ref{lpl}), completely cancel so that any nonzero measurement of
this value in the experiments will provide essential evidence of new
physics beyond SM. The combined normal polarization, $P_N^-+P_N^+$,
is zero beyond the SM, since $P_N^-$ and $P_N^+$ receive
contribution only from $Im(AQ)$ term with opposite sign.

A last note before going into details of numerical analysis is in
order. In the expressions of the lepton polarizations we note that
they all depend on $s$ and the new Wilson coefficients. Because of
experimental difficulties of studying the polarizations of each
lepton depending on both quantities, it would be better to eliminate
the dependence of the lepton polarizations on $s$, by considering
the averaged forms over the allowed kinematical region. The averaged
lepton polarizations are defined by \bea \label{av}
    \lla P_i \rra = \frac{\ds \int_{(2 m_\ell/m_{B_c})^2}^{(1-m_{D_s}/m_{B_c})^2}
                        P_i \frac{d{\cal B}}{ds} ds}
            {\ds \int_{(2 m_\ell/m_{B_c})^2}^{(1-m_{D_s}/m_{B_c})^2}
                     \frac{d{\cal B}}{ds} ds}~.
\eea
\section{Numerical analysis and discussion}
Before going on our numerical analysis of the branching ratios and the averaged polarization asymmetries
 $<P^-_L>$, $<P^-_T>$
and $<P^-_N>$ of $\ell^-$ for the \BcDll decays with $\ell =\mu , \tau $ as well as
the lepton-antilepton combined asymmetries $<P^-_L+P^+_L>$ and $<P^-_T-P^+_T>$,
let us first introduce the input parameters used in this work:
\begin{eqnarray}
& & m_{B_c} =6.50 \, GeV, \, \, m_{D_s}=1.968 \,GeV, \,\, m_b =4.8 \, GeV,
        \,\,m_{\mu} =0.105 \, GeV, \,\,m_{\tau} =1.77 \, GeV, \nnb \\
& &  |V_{tb} V^*_{ts}|=0.0385, \, \,\, \, \alpha^{-1}=129, \, \,G_F=1.17 \times 10^{-5}\, GeV^{-2}, \nnb \\
& &  \tau_{B_{c}}=0.46 \times 10^{-12} \, s, \,\, C^{eff}_7=-0.313, \,\,C^{eff}_9=4.344\,\,,C_{10}=-4.624
\end{eqnarray}
%
%

Details of the values of the individual Wilson coefficients in the
SM at $\mu \sim m_b $ scale can be found in \cite {uoyilmaz}.

The given value of $C^{eff}_9 $ corresponds only to the
short-distance contributions, but we know that $C^{eff}_9$ also
receives long-distance contributions due to conversion of the real
$\bar{c}c$ into lepton pair $\ell^+ \ell^-$, and they are usually
absorbed into a redefinition of the short-distance Wilson
coefficients:
\begin{eqnarray}
C_9^{eff}(\mu)=C_9(\mu)+ Y(\mu)\,\, ,
\label{C9efftot}
\end{eqnarray}
where
\begin{eqnarray}
\label{EqY}
Y(\mu)&=& Y_{reson}+ h(y,s) [ 3 C_1(\mu) + C_2(\mu) +
3 C_3(\mu) + C_4(\mu) + 3 C_5(\mu) + C_6(\mu)] \nonumber \\&-&
\frac{1}{2} h(1, s) \left( 4 C_3(\mu) + 4 C_4(\mu)
+ 3 C_5(\mu) + C_6(\mu) \right)\nnb \\
&- &  \frac{1}{2} h(0,  s) \left[ C_3(\mu) + 3 C_4(\mu) \right]
\\&+& \frac{2}{9} \left( 3 C_3(\mu) + C_4(\mu) + 3 C_5(\mu) +
C_6(\mu) \right) \nonumber \,\, ,
\end{eqnarray}
and $y=m_c/m_b$, and the functions $h(y,s)$ arises from the one loop
contributions of the four quark operators $O_1$,...,$O_6$ explicit
forms of which can be found in \cite{Buras}- \cite {Misiak2}.
Parametrization of the resonance $\bar{c}c$ contribution,
$Y_{reson}(s)$, given in Eq.(\ref{EqY}) can be done by using a
Breit-Wigner shape with normalizations fixed by data given in
\cite{Ali2}
\bea
Y_{reson}(s)&=&-\frac{3}{\alpha^2_{em}}\kappa \sum_{V_i=\psi_i}
\frac{\pi \Gamma(V_i\rightarrow \ell^+
\ell^-)m_{V_i}}{s m^2_B-m_{V_i}+i m_{V_i}
\Gamma_{V_i}} \nonumber \\
&\times & [ (3 C_1(\mu) + C_2(\mu) + 3 C_3(\mu) + C_4(\mu) + 3
C_5(\mu) + C_6(\mu))]\, ,
 \label{Yresx}
\eea
where the phenomenological parameter $\kappa$ is usually taken as
$\sim 2.3$.

The new Wilson coefficients are the free parameters in this work,
but  it is possible to establish ranges out of experimentally measured branching ratios of the
semileptonic rare B-meson decays
\bea
    BR (B \rar K \, \ell^+ \ell^-) & = & (4.8^{+1.0}_{-0.9}\pm 0.3) \times 10^{-7} \, \, ,\nnb \\
    BR (B \rar K^* \, \mu^+ \mu^-) & = & (1.27 ^{+0.76}_{-0.61}\pm
                    0)\times 10^{-6}\, \, ,\nnb \eea
reported  by Belle and Babar collaborations \cite{Belle}-\cite
{Babar}, also the upper bound of pure leptonic rare B-decays in the
$B^0 \rar \mu^+ \mu^-$ mode \cite{CDFII}: \bea
    BR ( B^0 \rar  \mu^+ \mu^-) & < 1.5 \times 10^{-7}  \, \, .\nnb
\eea

All new Wilson coefficients are taken as real and varying in the
region $-4\leq C_X\leq 4$, compliant to this upper limit and the
above mentioned measurements of the branching ratios for the
semileptonic rare B-decays.

The new Wilson coefficients in Eq.(\ref{effH}), the helicity-flipped
counter-parts of the SM operators, $C_{RL}$ and $C_{RR}$, vanish in
all models with minimal flavor violation in the limit $m_s \rar 0$.
However, in some MSSM scenarios there exist finite contributions
from these vector operators even for a vanishing s-quark mass. In
addition, scalar type interactions can also contribute through the
neutral Higgs diagrams, multi-Higgs doublet models and MSSM, for
some regions of the parameter spaces of the related models. In
literature there are some studies to establish ranges out of
constraints under various precision measurements for these
coefficients (see e.g. \cite{HuangWu}) and our choice for the range
of the new Wilson coefficients are in agreement with these
calculations.

To make some numerical predictions, we also need the explicit forms
of the form factors $F_+, F_-$ and $F_T$. Since there are two heavy
quarks the non-relativistic effects and also the additional
contributions from hard interactions might give useful information.
In this work we have not considered these effects and used the
results of \cite{Faessler}, calculated in a relativistic constituent
quark model in which
 $q^2$ dependencies of the form factors
are given as
\begin{eqnarray}
F(q^2) = \frac{F(0)}{\Big( 1-a s + b s^2 \Big)^2}~, \nnb
\end{eqnarray}
where values of parameters $F(0)$, $a_F$ and $b_F$ for the $B_c \rar
D_s$ decay are listed in Table 1.
\begin{table}[h]
\renewcommand{\arraystretch}{1.5}
\addtolength{\arraycolsep}{3pt}
$$
\begin{array}{|l|ccc|}
\hline
& F(0) & a & b \\ \hline
F_+ & \phantom{-}0.186   & 2.48 & \phantom{-} 1.62 \\
F_- &\phantom{-} -0.190  & 2.44 & \phantom{-} 1.54\\
F_T & \phantom{-}0.275   & 2.40 & \phantom{-}1.49\\
\hline
\end{array}
$$
\caption{$B_c$ meson decay form factors in a relativistic constituent quark model without impulse
approximation .}
\renewcommand{\arraystretch}{1}
\addtolength{\arraycolsep}{-3pt}
\end{table}

Before the discussion of the results of our analysis given in a series of figures, we give our SM predictions
for the longitudinal, transverse and the normal components of the
lepton polarizations for \BcDll decay for $\mu$ ($\tau$) channel
for reference: \bea
<P^-_{L}>  & =  & -0.8457 \, (-0.1774) \, ,\nnb \\
<P^-_{T}>  & =  & -0.0948 \, (-0.6189) \, ,\nnb \\
<P^-_{N}>  & =  & 0 \, (0) \, .\nnb
\eea

\begin{figure}
\centering
\includegraphics[width=5in]{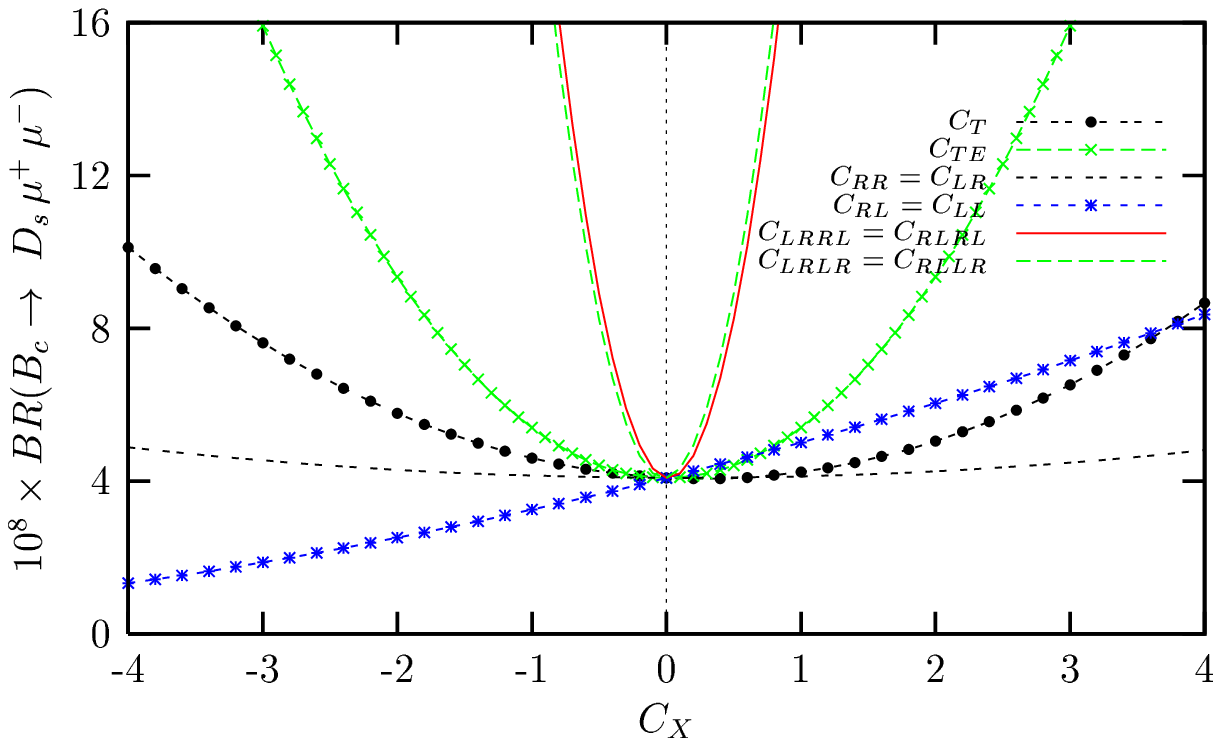}
\caption{The dependence of the integrated branching ratio for the
$B_c \rar D_s \, \mu^+ \mu^-$ decay on the new Wilson
coefficients. \label{f1}}
\end{figure}

\begin{figure}
\centering
\includegraphics[width=5in]{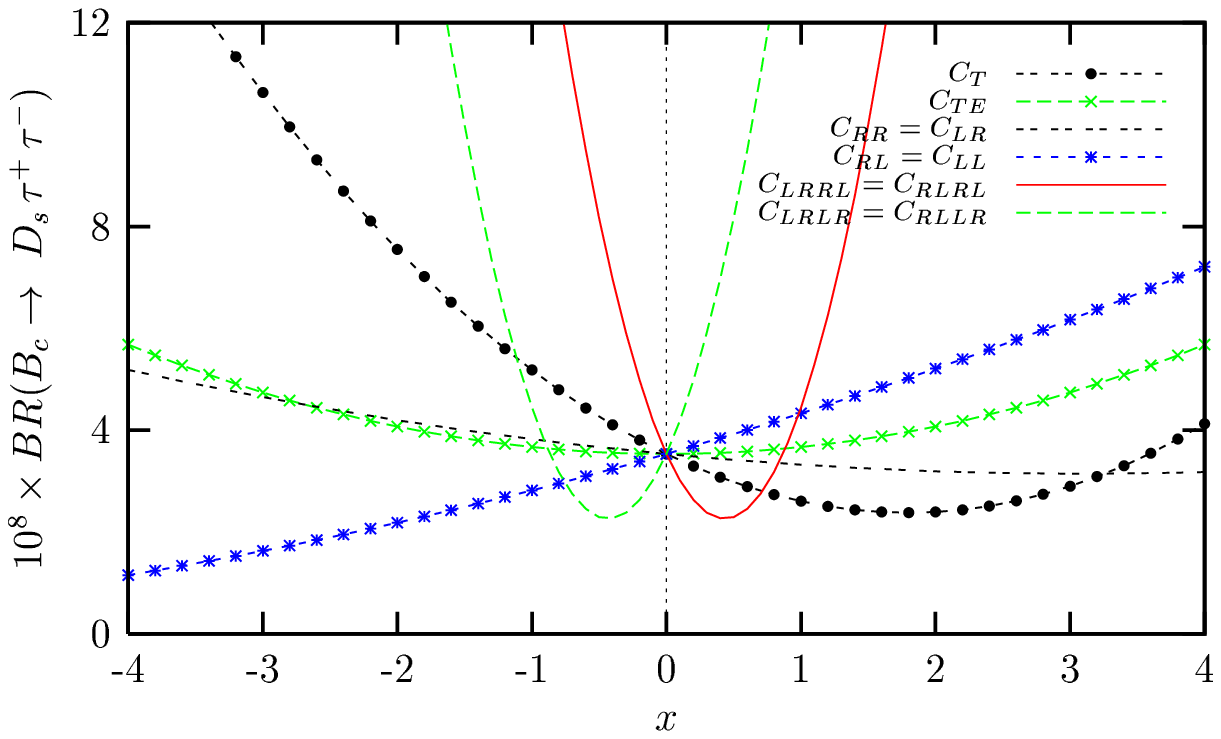}
\caption{The dependence of the integrated branching ratio for the
$B_c \rar D_s \, \tau^+ \tau^-$ decay on the new Wilson
coefficients. \label{f2}}
\end{figure}

In Figs. (\ref{f1}) and (\ref{f2}), we give the dependence of the
integrated branching ratio (BR) on the new Wilson coefficients for
the \BcDllmu and \BcDlltau decays, respectively. From these figures
we see that BR depends strongly on the scalar and tensor
interactions and weakly on the vector interactions. It is also clear
from these figures that dependence of the BR on the new Wilson
coefficients is symmetric with respect to the zero point for the
muon final state, but such a symmetry is not observed for the tau
final state for the tensor interactions. Another remark is that, for
muon case $C_{TE}$ is dominant while for tau case $C_T$ becomes more
dominant.

\begin{figure}
\centering
\includegraphics[width=5in]{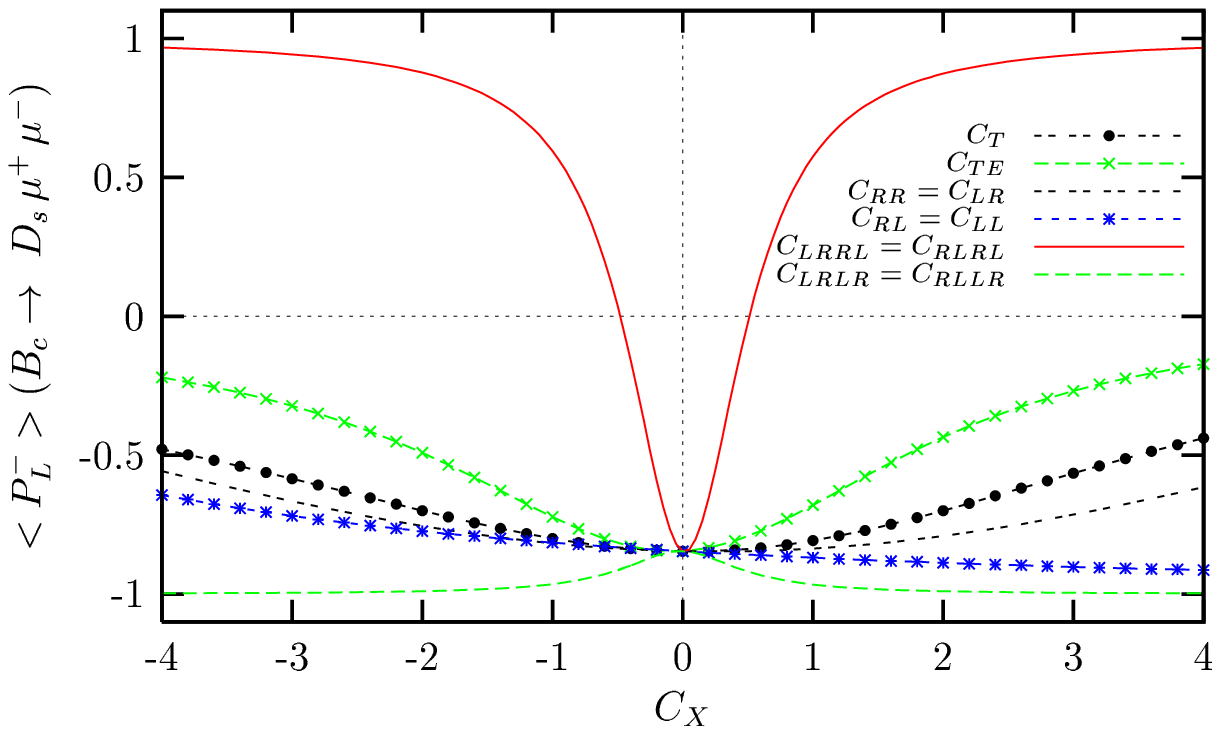}
\caption{The dependence of the averaged longitudinal polarization
$<P^-_L>$ of $\ell^-$ for the $B_c \rar D_s \, \mu^+ \mu^-$ decay
on the new Wilson coefficients. \label{f3}}
\end{figure}

\begin{figure}
\centering
\includegraphics[width=5in]{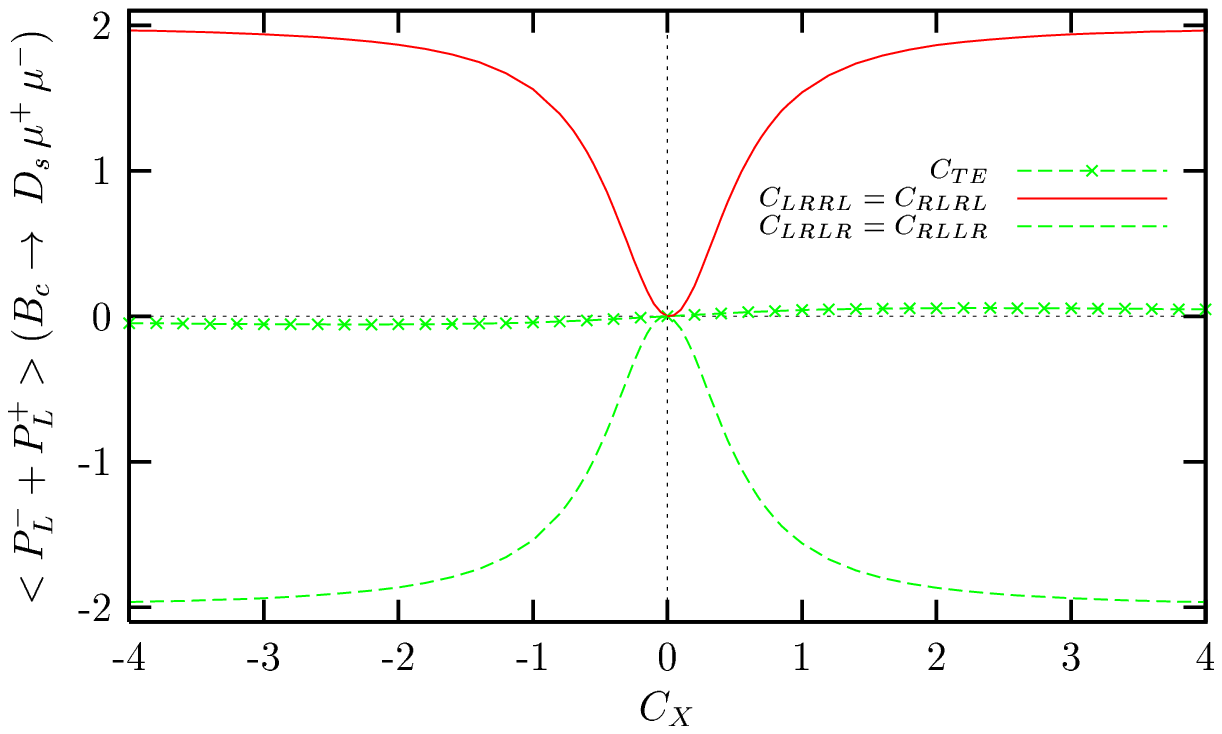}
\caption{The dependence of the combined averaged longitudinal lepton
polarization \, \, \, $<P^-_L+P^+_L>$ for the $B_c \rar D_s \,
\mu^+ \mu^-$  decay on the new Wilson coefficients.\label{f4}}
\end{figure}

In Figs. (\ref{f3}) and (\ref{f4}), we present the dependence of
averaged longitudinal polarization $<P_L^->$ of $\ell^-$ and the
combined averaged $<P_L^- + P_L^+ >$ for \BcDllmu decay on the new
Wilson coefficients. It is observed that the dominant contribution
for $<P_L^- >$ comes from the scalar interactions of the type
$C_{LRRL}$ and $C_{RLRL}$ which are identical and symmetric with
respect to $C_X=0$ while the combined averaged $<P_L^- + P_L^+ >$ is
sensitive to that of scalar type interactions only. It is a
well-known fact that vector type interactions are canceled when the
longitudinal polarization asymmetry of the lepton and antilepton is
considered together. This is the reason $<P_L^- + P_L^+>$ does not
exhibit any vector type dependence in Fig. (\ref{f4}). It is also
interesting to note that $<P_L^- + P_L^+ >$ is positive for
$C_{LRRL}$ and $C_{CRLRL}$ and negative for other scalar
contributions. In addition they are symmetric with respect to
$C_X=0$.

\begin{figure}
\centering
\includegraphics[width=5in]{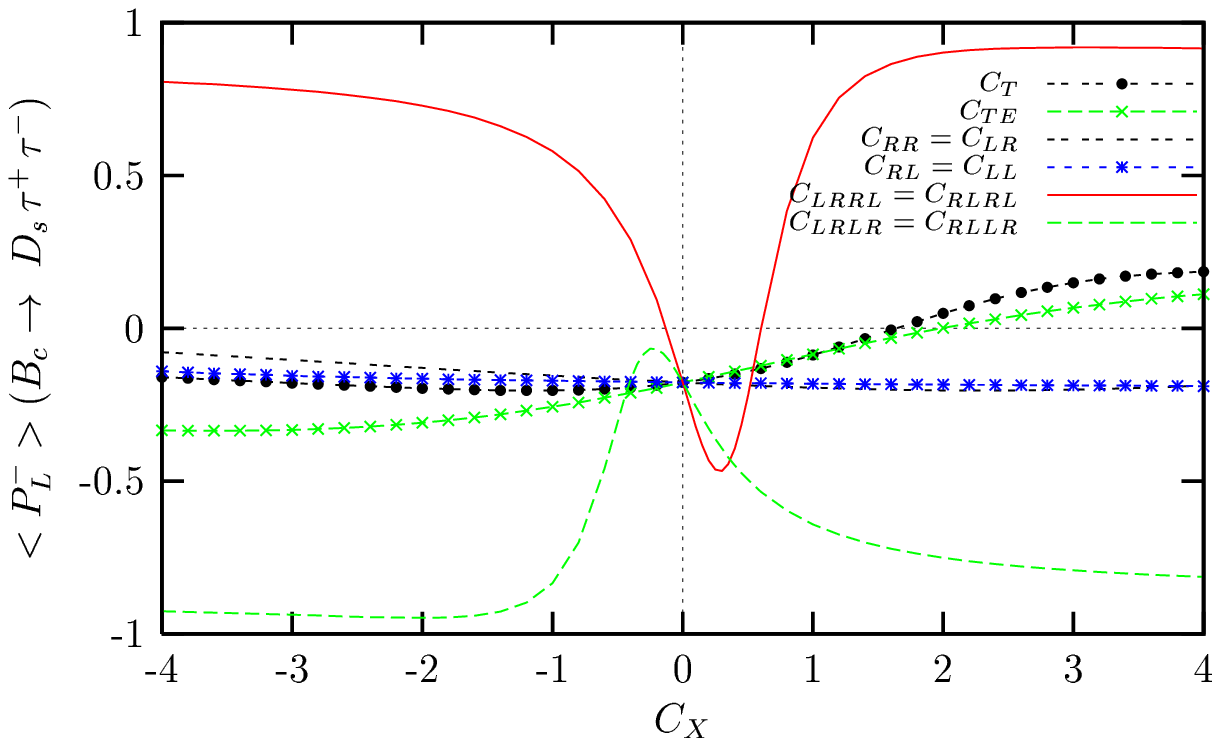}
\caption{The same as Fig. (\ref{f3}), but for the $B_c \rar
 D_s \, \tau^+ \tau^-$ decay. \label{f5}}
\end{figure}

\begin{figure}
\centering
\includegraphics[width=5in]{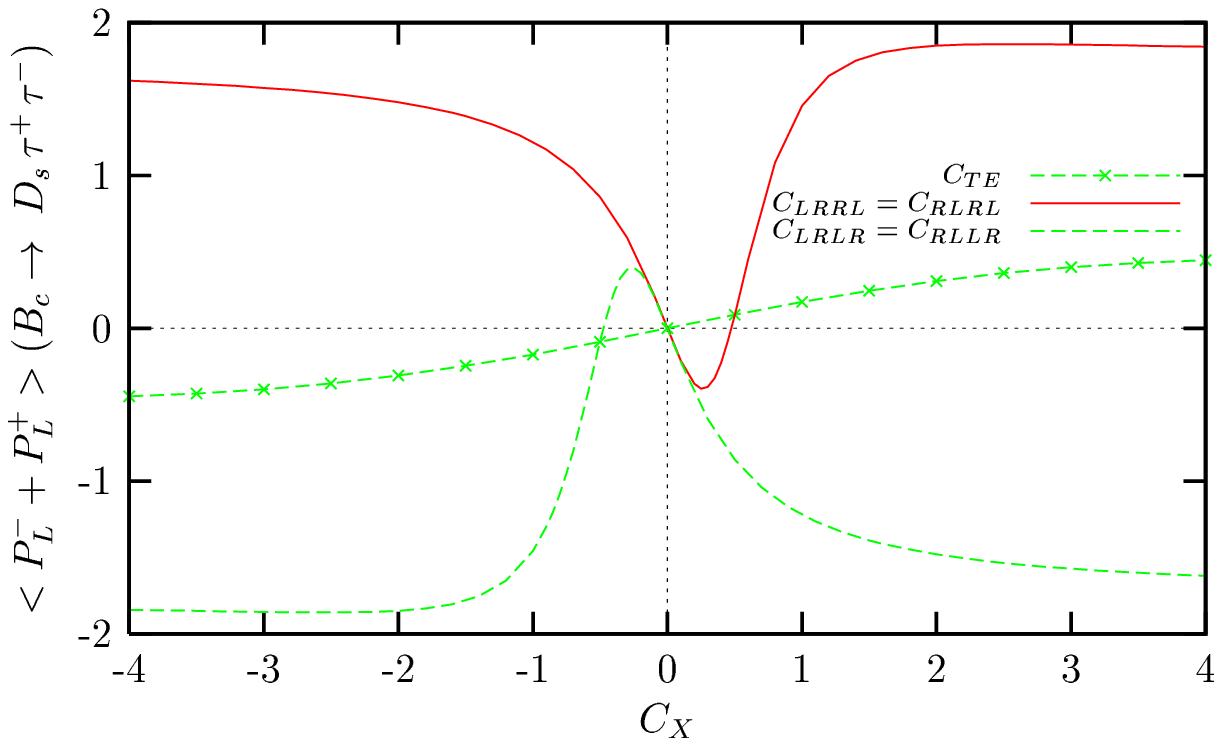}
\caption{The same as Fig. (\ref{f4}), but for the $B_c \rar
 D_s \, \tau^+ \tau^-$ decay. \label{f6}}
\end{figure}

Figures (\ref{f5}) and (\ref{f6}) are the same as Figs. (\ref{f3})
and (\ref{f4}), but for \BcDlltau. $<P_L^- >$ strongly depends on
scalar type interactions and also sensitive to tensor type
interactions. In case of vector interactions, they are nearly
identical for $C_X>0$ and keeping their SM values. As in the muon
case, $<P_L^- + P_L^+ >$ depends on scalar interactions and $C_{TE}$
tensor interactions. $C_{TE}$ effect is comparably higher than that
of muon case and it is negative (positive) for $C_X<0$ ($C_X>0$).
For $<P_L^- + P_L^+ >$ in both $\mu$ and $\tau$ decays, there is no
SM effect and any nonzero experimental results of $<P_L^- + P_L^+ >$
will be the evidence of new physics.

\begin{figure}
\centering
\includegraphics[width=5in]{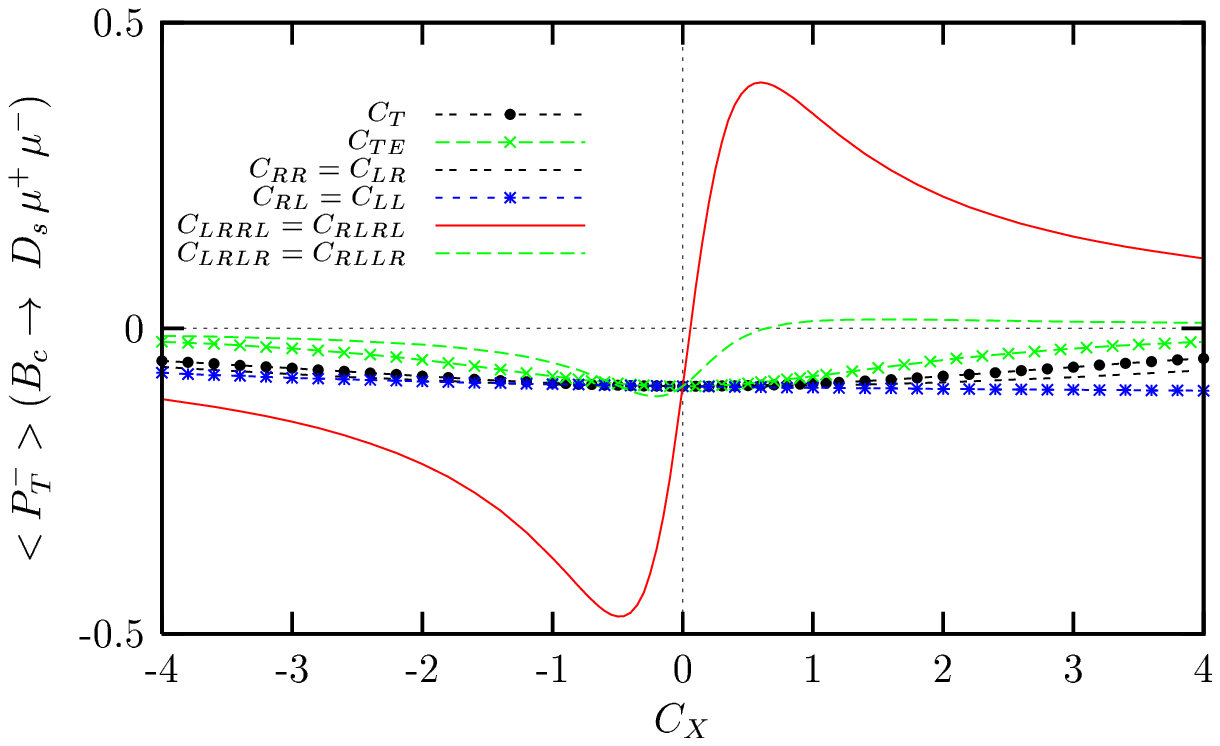}
\caption{The dependence of the averaged transverse polarization
$<P^-_T>$ of $\ell^-$ for the $B_c \rar D_s \, \mu^+ \mu^-$
decay on the new Wilson coefficients. \label{f7}}
\end{figure}

\begin{figure}
\centering
\includegraphics[width=5in]{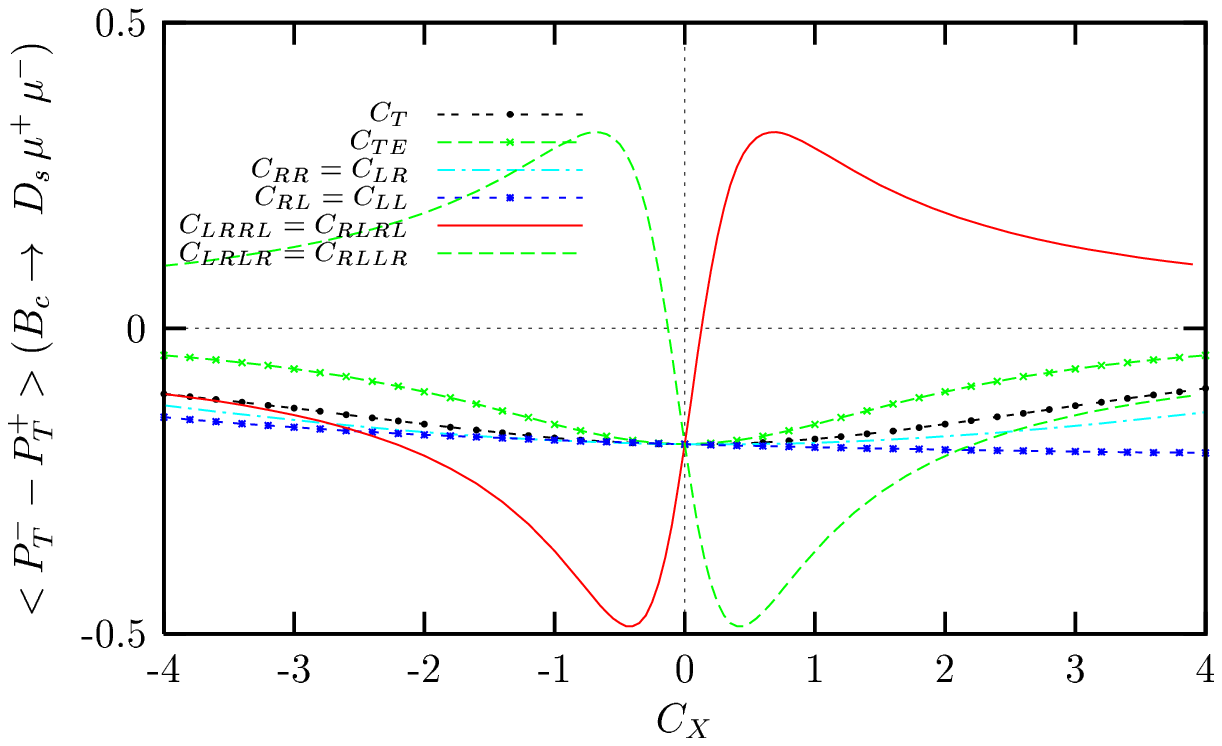}
\caption{The dependence of the combined averaged transverse lepton
polarization \, \, \, $<P^-_T-P^+_T>$ for the $B_c \rar
D_s \gamma \, \mu^+ \mu^-$  decay on the new Wilson
coefficients.\label{f8}}
\end{figure}

In Figs. (\ref{f7}) and (\ref{f8}), we present the dependence of
averaged transverse polarization $<P_T^->$ of $\ell^-$ and the
combined averaged $<P_T^- - P_T^+ >$ for \BcDll decay on the new Wilson coefficients.
The significant effect of scalar $C_{LRRL}$ and $C_{RLRL}$ can be seen from Fig. (\ref{f7}).
Here, $<P_T^->$ is negative (positive) for $C_X\simlt 0$ ($C_X\simgt 0$). As a last remark,
all other contributions are negative
except $C_{LRLR}$ and $C_{RLLR}$ scalar terms for $C_X\simgt 0.8$.
Unlike the $<P_T^->$, $C_{LRLR}$ and $C_{RLLR}$ make greater contribution to $<P_T^- - P_T^+ >$ as seen in Fig. (\ref{f8}).
Considering Figs. (\ref{f7}) and (\ref{f8}) together, determination of the sign and magnitude of the scalar observables
can also give useful information about the existence of new physics.

\begin{figure}
\centering
\includegraphics[width=5in]{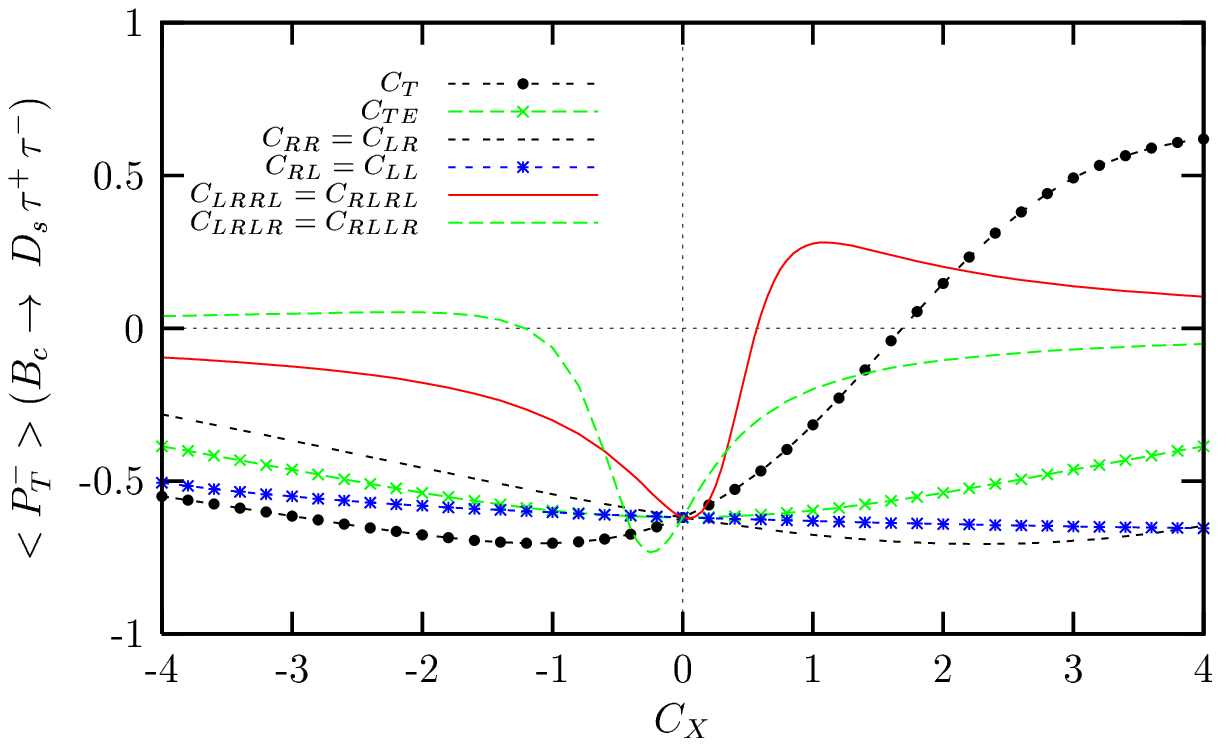}
\caption{The same as Fig. (\ref{f7}), but for the $B_c \rar
 D_s \, \tau^+ \tau^-$ decay. \label{f9}}
\end{figure}

\begin{figure}
\centering
\includegraphics[width=5in]{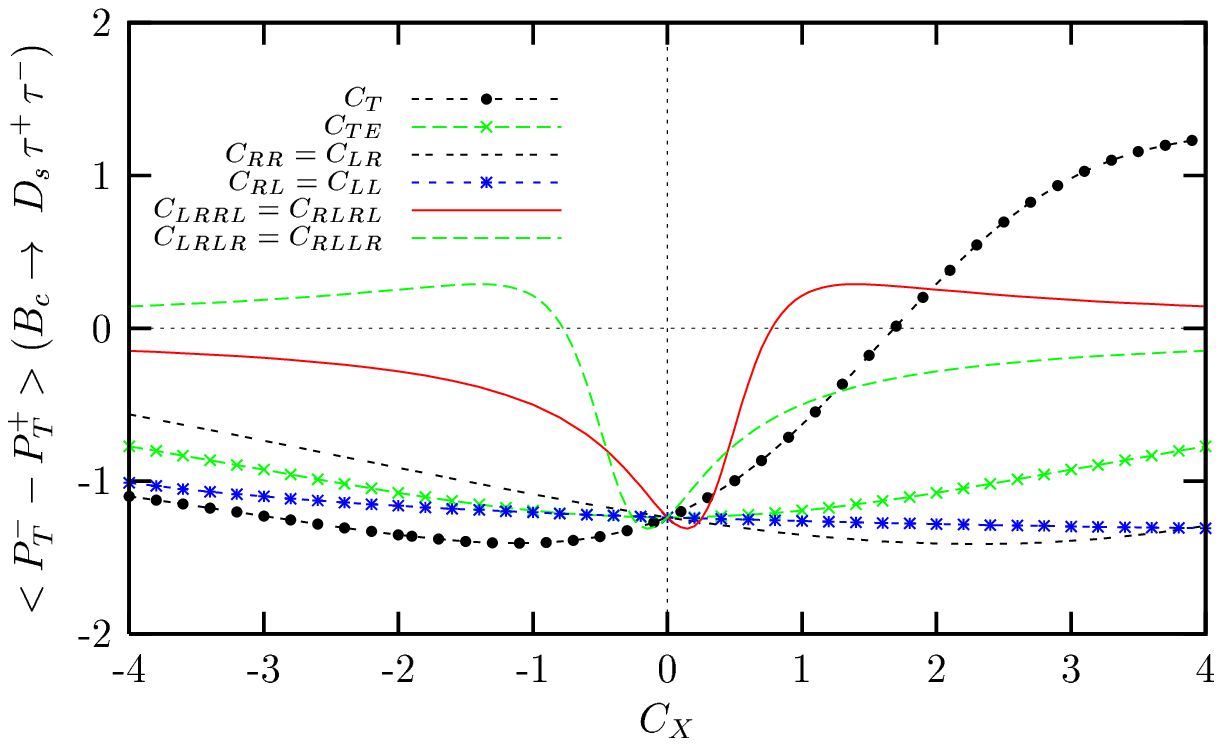}
\caption{The same as Fig. (\ref{f8}), but for the $B_c \rar
 D_s \, \tau^+ \tau^-$ decay.\label{f10}}
\end{figure}

Figures (\ref{f9}) and (\ref{f10}) are the same as Figs. (\ref{f7})
and (\ref{f8}), but for \BcDlltau. In both figures, the effects of
scalar contributions are clear. Considering $<P_T^- - P_T^+ >$ it
can be noted that values of obtained data are two times that of
$<P_T^->$ values.

\begin{figure}
\centering
\includegraphics[width=5in]{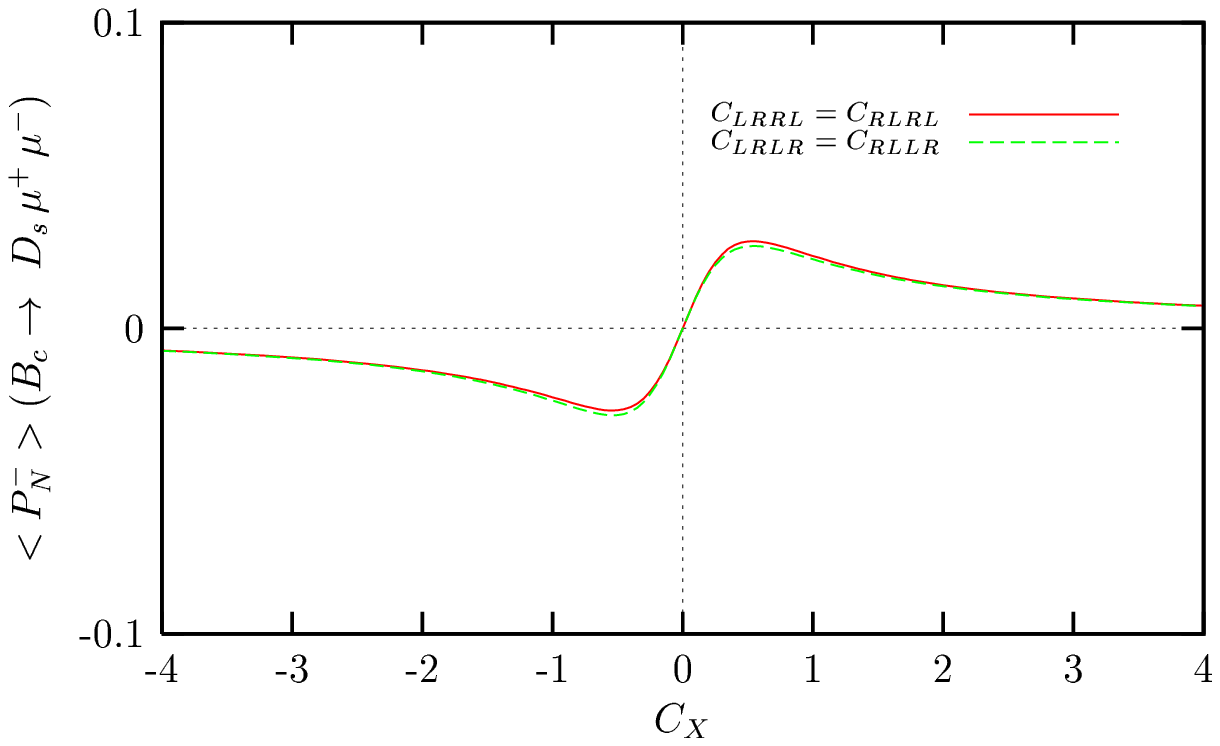}
\caption{The dependence of the averaged normal polarization
$<P^-_N>$ of $\ell^-$ for the $B_c \rar D_s \, \mu^+ \mu^-$
decay on the new Wilson coefficients.\label{f11}}
\end{figure}

\begin{figure}
\centering
\includegraphics[width=5in]{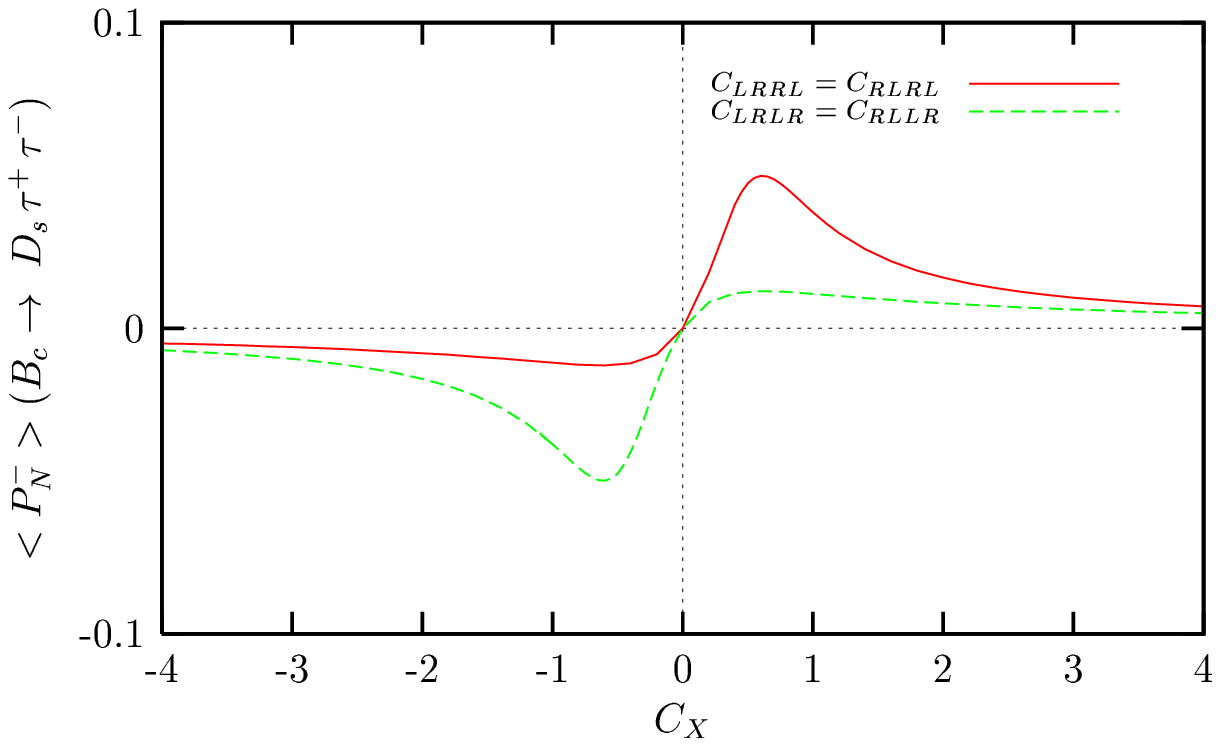}
\caption{The same as Fig.(\ref{f11}), but for the $B_c \rar D_s \, \tau^+ \tau^-$  decay.\label{f13}}
\end{figure}

Figures (\ref{f11}) and (\ref{f13}), we present the dependence of
averaged normal polarization $<P_N^->$ of $\ell^-$ for \BcDllmu and \BcDlltau
decays on the new Wilson coefficients.
Normal components of polarization of both channel depend only on scalar type interactions.
As discussed before, SM contribution of normal polarization is zero so any measurable values should come
from new physics. These values will also give promising information on sign and magnitude of $<P_N^->$ because $<P_N^->$
is negative (positive) for $C_X<0 (C_X>0)$.

In conclusion, using the general, model independent form of the
effective Hamiltonian, we present the most general analysis of the
lepton polarization asymmetries in the rare \BcDll decay. The
dependence of the longitudinal, transversal and normal polarization
asymmetries of $\ell^-$ and their combined asymmetries on the new
Wilson coefficients are studied. The lepton polarization asymmetries
are very sensitive to the existence of the scalar type interactions
and in some cases tensor type interactions worth to be considered.
Individually, $C_{LRRL}$ and $C_{RLRL}$ play a significant role
throughout this work. In all types of analysis the following terms
are found identical: $C_{RR}=C_{LR}$, $C_{RL}=C_{LL}$,
$C_{LRRL}=C_{RLRL}$ and $C_{LRLR}=C_{RLLR}$. Moreover, in the most
cases polarization effects change their signs  as the new Wilson
coefficients vary in the region of interest, which is useful to
determine the sign in addition magnitude of new physics effect. A
last note on combined asymmetries, a well known SM result that
$<P_L^- + P_L^+> =0$, $<P_N^- + P_N^+>=0$ and $<P_T^- - P_T^+>
\simeq 0$ in the limit $m_\ell \rar 0$. Therefore any deviation from
these relations and determination of the sign of polarization is
decisive and effective tool in searching new physics beyond the
SM. \\

{\large \bf Acknowledgments}\\
\\
The author would like to thank G. Turan and A. Ozpineci for valuable
contributions and discussions.  This work was partially supported by
Mersin University under Grant No: BAP-FEF-FB (UOY) 2006-3.

\newpage
\renewcommand{\topfraction}{.99}
\renewcommand{\bottomfraction}{.99}
\renewcommand{\textfraction}{.01}
\renewcommand{\floatpagefraction}{.99}

\clearpage
\clearpage
\clearpage
\clearpage
\clearpage
%
%
%

\end{document}